\documentclass[aps,prl,reprint]{revtex4-1}
\usepackage{blindtext}
\usepackage{graphicx}
\usepackage{caption}
\usepackage{xcolor}
\usepackage{adjustbox}

\begin{document}
\title{Relativistic reference frame PIC simulations for electron beam dynamics with meter-scale propagation inside plasma and under external fields}
%\title{\textcolor{black}{Electron beam dynamics along meter-scale propagation through plasma pre-buncher and under external field via Relativistic reference frame PIC simulations}}
\author{Driss Oumbarek Espinos$^{1}$}
\email{doumbare@post.kek.jp}
\author{Alexei Zhidkov$^{2,3}$}
\author{\\Alexandre Rondepierre$^{2,3}$}
\author{Masafumi Tawada$^{1}$}
\author{Mika Masuzawa$^{1}$}
\affiliation{$^{1}$KEK, 1-1 Oho, Tsukuba, Ibaraki 305-0801, Japan.}
\affiliation{$^{2}$ Institute of Scientific and Industrial Research (SANKEN), Osaka University, 8-1 Mihogaoka, Ibaraki, 565-0871, Osaka, Japan.}
\affiliation{$^{3}$ Laser Accelerator R$\&$D, Innovative Light Sources Division, RIKEN SPring-8 Center, 1-1-1, Kouto, Sayo-cho, Sayo-gun, Hyogo, 679-5148, Osaka, Japan.}

\begin{abstract}
Particle in cell simulations are widely used in most fields of physics to investigate known and new phenomena which cannot be directly observed or measured yet. However, the computational and time resources needed for PICs make them impractical when high resolution and long time/distance simulations are required. In this work, we present a new PIC simulation code that takes advantage of the use of a relativistic reference frame and consequent time dilation and length contraction.
These properties make a simulation capable of long (meter length) and high resolution simulations without the need of supercomputers. This new code is a step forward with regards to the previous tries enabling complex multiple body situations without additional filtering and smoothing of fields and currents.
The usefulness of the relativistic frame PIC code is displayed by simulating electron beam bunching obtained in long undulator propagation and also the potential as a beam "buncher" of 10s of cm long low density plasmas.     
\end{abstract}

\maketitle

\section{Introduction}

Essential advantage in elaboration of plasma accelerators \cite{joshi2020perspectives} and plasma beam optics \cite{doss2019laser,pathak2023focusing} stimulates the developing of other plasma elements able to reduce the size and cost of conventional devices.\textcolor{black}{ One of such instruments is the undulator \cite{luchini1990undulators}, especially for XFEL generation \cite{rosenzweig2020ultra}, which needs of 10s of meters of it to obtain the desired radiation. Adding plasma elements in its construction or even developing a full plasma undulator \cite{kuzelev1995basics,rykovanov2015plasma} may drastically improve the characteristics of these radiation devices operating at far shorter periods with the same K-number \cite{luchini1990undulators}. Both of these cases requires thorough numerical investigation to understand the way how to develop, as the analytical approaches \cite{kuzelev1995basics} are surely not enough, as one needs to carefully comprehend the intricate interactions between the electron beam particles and everything else.}

\textcolor{black}{ 
There are several numerical softwares for simulating the characteristics of vacuum undulators \cite{Genesis14,tanaka2015simplex,floettmann2007space}. However, simulation of complex setups like the plasma interaction with particle beams in the presence of an external fields can be performed only by particle-in-cell methods. }

Since its inception in the 1960s \cite{langdon2014evolution, birdsall2018plasma} Particle in cell (PIC) simulations have become an indispensable tool to study the underlying physics of complex phenomena which cannot be accessed in any other way. PIC permits to model the kinetics of a grand ensemble of particles interacting with each other and with electromagnetic fields in a self-consistent manner. In a PIC code, A finite number of macroparticles, i.e., groups of real particles that conserve the mass-to-charge ratio, sample the system phase-space well enough to achieve a good results without the need of a large number of real particles. Such codes offer a great flexibility, thus, virtually any system could be simulated in an 'ab initio' fashion. However, two major factors limit the applications of PIC. Firstly, for highly complex systems, a lot of different types of phenomena can occur at the same time (ionization \cite{zhidkov1998hybrid}, collision \cite{oumbarek2018langevin, sentoku2008numerical}, acceleration \cite{kruer1988physics}, etc), which requires careful planning of the code and testing to make sure that everything occurs as expected. Regarding the code, the different numerical methods have distinct drawbacks (accuracy \cite{nikitin2006third, van1988leap,pathak2022effect}, charge conservation \cite{esirkepov2001exact}, border conditions \cite{tajima2018computational}, etc). On top of that, testing that everything is correct needs of the creation of benchmark cases where the result should be understood before hand, which, in some cases is a challenge on itself.    
Secondly, the sheer number of macroparticles and field points defined in a fine spatial grid (in order to resolve the smallest elements) may demand huge amounts of computational resources for a 6D phase-space. In some instances, the phase-space can be reduced by just simulating a 2D geometry or taking advantage of symmetries like with the quasi-3D cylindrical codes \textcolor{black}{\cite{lifschitz2009particle, lehe2016spectral}}. However, such dimensional reduction can be only used in specific problems and even then, energy conservation, numerical heating \cite{birdsall2018plasma} or the physics of the problem may be affected, especially for long simulations. So, the use of such tools require a previous study to ascertain if they are appropriate and to what extent. 
Even using the dimensional reduction, problems that require long propagation distances ($>$cm) are not practical to do due to the usual required resolution and simulation box size. For example, in principle, the simple propagation of a relativistic electron beam along a magnetic planar undulator to generate free electron laser (FEL) radiation \cite{madey1971stimulated}, becomes a difficult task as it needs meters of propagation with a spatial resolution smaller than the electron beam radiation wavelength. A quick estimation tells us that for a radiation wavelength of $\lambda_r$ = 200 $nm$, a proper simulation will require a spatial resolution of $\Delta x$=$\frac{\lambda_r}{16}$ = 12.5 $nm$, thus, a meter of propagation needs $8\times10^{7}$ steps, however, the bunching and initialization of coherent radiation for FEL \cite{elias1976observation} requires of typically $\approx$1-3 $m$, i.e., $\approx 8-24\times10^{7}$ steps. Simulating such number of steps for $\approx 10^{8}$ particles requires of huge resources including simulation time. In addition, the meter size simulation box cannot be calculated directly as it will require a prohibitive amount of RAM ($\approx 1$ PB) to be stored during the simulation. While the later issue can be solved by using a 'moving window' \cite{tajima2018computational} simulation box, the former is not trivial.

In the particular case of FEL generation, one might think that the already existing codes, e.g., GENESIS \cite{Genesis14}, SIMPLEX \cite{tanaka2015simplex}, ASTRA \cite{floettmann2007space}, are enough as they allow to predict the experimental FEL performance even though they are not 'ab initio' and use multiple approximations. While this is true if one only cares about such usage, they are not enough to properly study the beam dynamics and most importantly the three element interaction previously alluded to, e.g., between the undulator field, electron beam and a plasma. The addition of another element like a plasma to this process cannot be done with the aforementioned codes and it is of great interest for the future undulators as cm size plasmas show great promise as an electron beam early buncher. Furthermore, even the electron beam propagation inside 10s of cm long plasmas cannot also be fully done in a reasonable time and set of resources, and such simulations are of capital importance for the research of plasma optics and plasma undulators, which require both long propagation and enough resolution for the radiation emitted.

\textcolor{black}{With the objective of understanding the use of a plasma optic to improve the beam characteristics to increase the efficiency of the FEL radiation generation and exploring the real use possibilities different PIC existing solutions were explored, however, none adapted to such research was discovered. Consequently, this novel research required the development of appropriate simulation tools.}

\textcolor{black}{In this work, an extension of the PIC code FPlaser \cite{zhidkov2004effects} to approach such problematic (arbitrary plasma-beam-external field interactions) is presented and used for our research , i.e., in the context of the interaction between a relativistic electron beam, a planar undulator and a plasma. In addition, the first preliminary results of the plasma element effect on electron beams showing the potential are presented.}
The new code is based on the use of a booster reference frame that moves, in the electron beam propagation direction, at a constant relativistic speed, i.e., $\gamma_{R}>1$. In this relativistic reference frame, following the relativity equations, the meter (cm) size undulator (plasma) is reduced to $\approx cm$ ($\approx mm$) size. On the other hand, the $\mu m$ Gaussian electron beam elongates to only $mm$ size. 
These changes in distance make suddenly possible to use PIC to calculate the electron beam propagation even for meters without the need of excessive computer resources. However, being in a relativistic reference frame (RRF) instead of a laboratory reference frame (LRF) raises a lot of questions about the way to do and understand the simulation results of the complex 3-way interaction that require of careful research. \\

Previously, such RRF has been tried in the code WARP \cite{friedman1990warp, vay2007noninvariance} for external injection LPA \cite{vay2011effects}, interaction between a ultra-relativistic proton beam and simple uniform magnetic field \cite{vay2008simulation}. However, multiple instabilities related to the back-scattering radiation and at the plasma column that can end up introducing large quantities of noise is a recurring problem. This issues limit the $\gamma_{R}$ possible for each simulation case. Nevertheless, great efforts are being carried to filter the instabilities manually using different methods and modified pushers \cite{vay2011numerical, vay2008simulation}, thus, allowing in some cases high $\gamma_{R}$. In addition, a 2D FEL simulation was tried (low current electron beam in a periodic magnetic field) \cite{fawley2011full} in which no moving window was used and the electron beam was selected so the space-charge was not prevalent and no interaction with the window borders appeared. However, while nothing inherently wrong was identified, the simulation could not properly reproduce the SASE (self-amplified spontaneous emission) phenomena and even with a prepared artificial seed the results differed from the expected results by a factor from 2 to 4 times. Some of the issues encountered where due to the choice of $\gamma$.
Around the same time, and following WARP's, the OSIRIS code \cite{fonseca2002osiris} also implemented a RRF for its PIC simulations. In this case, again, it has been mostly oriented for self-injection, injection and external guiding and LPA simulations \cite{martins2010exploring, martins2010numerical} and to simulate electron beam radiation during betatron oscillations in a ion-channel like electric field \cite{davoine2018ion}. Nevertheless, the same instability problems are found. The simulations are limited by the quantity of backwards radiation and its wavelength \cite{martins2010exploring, martins2010numerical}, and in applicable cases they are filtered. Moreover, field and plasma current smoothing are used to avoid instabilities growth \cite{martins2010exploring}.

While not reinventing the PIC methods (e.g., leap-frog scheme, Poisson, etc), these previous works are steps forward in the right direction to generalize PICs methods to phenomena that requires of cm and meter sizes. However, as shown in the different previous works numerous problems of the existing solution limit the utility and applicable circumstances and are not appropriate for every situation. This work is another step forward to expand the PIC RRF to complex systems, including proper ion motion, transformation of beam distribution from LRF to RRF and simulation of high charge density electron beams. \textcolor{black}{And while not applicable to all cases, the here presented code has been conceived for the accurate simulation of the beam dynamics.}

\textcolor{black}{Numerical Cherenkov radiation (NCR), which is caused by the discretization of space and time in simulations deriving in an inaccurate speed of light can cause instabilities and non-physical phenomena \cite{vay2011numerical, pukhov2020x, godfrey2014suppressing, friedman1990second, godfrey2015improved, lehe2016elimination}.
%To reduce the possible NCR the more exact speed of light calculation following previous works by N. Pathak \cite{pathak2022effect} has been used.  Moreover, in the RRF where the beam is at rest ($\gamma_R = \gamma_B$) this problem completely disappears.}
\textcolor{black}{While in this work a laser is not taken into account it is important to remark that there is an essential difference between changing the reference frame for a particle beam (like is being done here) and a laser pulse. The RRF of the beam at rest ($\gamma_R = \gamma_B$) does not exist for a laser, always moving at the speed of light. In the LRF, the most important parameter for the interaction between a plasma and a laser is $\delta=\frac{N_e}{N_{cr}}$, with $N_e$ the plasma density and $N_{cr}=\frac{m \omega^2}{4\pi e^2}$ the laser critical density. If $\delta>1$ the laser cannot propagate (overdense plasma) and if $\delta<1$ the laser moves through the plasma (underdense plasma). Going to an RRF of $\gamma_R$, the previous quantities are transformed as: $N_{e,RRF}=\gamma_R N_{e,LRF}$ ; $m_{RRF}=\gamma_R m_{LRF}$ ; $\omega_{RRF}=\frac{\omega_{LRF}}{2\gamma_R}$. Therefore, the $\delta$ parameter becomes $\delta_{RRF}=4\gamma_{R}^{2}\delta_{LRF}$. For LWFA, the typical density of interest is $N_e\approx 10^{19} cm^{-3}$ and for a Ti-Sapphire laser pulse $N_{cr}=1.7x10^{21} cm^{-3}$, thus, in the RRF, the plasma becomes overdense at $\gamma_R\approx10$. It is well known from PIC simulations, that, even in the LRF, a laser pulse propagation into plasma with $N_e$ close to $N_{cr}$ is unstable resulting in filamentation and soliton formation \cite{zhidkov2007giant, sentoku1999bursts}. Meaning that, for a laser pulse the use of a RRF with higher gamma should result in some numerical instabilities related to the incorrect dispersion calculation, for example, NCR. 
Electron beams are free of this problem, and many instabilities vanish or affect only plasma after relaxation thus, not the beam dynamics.}
This work also goes deeper about the discussion regarding the extraction of a full picture of the system in the LRF from RRF data, and how it is limited by the particles own momentum and the time resolution. Nonetheless, as in the mentioned previous works \cite{martins2010exploring, martins2010numerical, vay2011numerical, vay2008simulation, vay2011effects, vay2007noninvariance}, the full transformation is not done here.}\\
%VENTUS (\textbf{V}ariable \textbf{E}nergy-frame \textbf{N}umerical \textbf{T}ool for \textbf{U}nderstanding of physic\textbf{S}) 
\textcolor{black}{In this work, first the basis for such PIC code \cite{EPHEMER} are explained, how the RRF is created and some 2D results to confirm the correctness of the simulation. Then, to present the promising usefulness of such PIC code and the concept and feasibility of the plasma element to improve the beam parameters, thus, increasing the Pierce Parameter \cite{bonifacio1984collective} (figure of merit for FEL) for later FEL radiation purposes, new results about the utilization of a low-density plasma as a tool to bunch the electron beam in cm lengths are shown, however, a full deep study will be made separately.}   
In Section 1, the general PIC scheme used and the implications of working in a relativistic frame will be presented. Section 2 deepens in the problematic of understanding the results in a RRF instead of a LRF. 
Then, in Section 3, the code is tested with a vacuum and inside undulator beam propagation simulations and the results are compared with the expected results.
Section 4 presents an overview of new results about the bunching imprinted into an electron beam during its propagation in a low density plasma, obtained thanks to the advantages offered by such code \cite{EPHEMER}. Before the conclusion, Section 5 swiftly shows the simulation of a complex system of a beam propagating inside a plasma inside an undulator.

%\section{Results}

\section{Relativistic booster frame}

The PIC code solves the electromagnetic equations via the Maxwell equations, constituting a system enough for the simulation:

\begin{equation}
\frac{\partial \textbf{E}}{\partial t} = c \nabla \times \textbf{B} - 4 \pi (\textbf{j}_{e}+\textbf{j}_{i})
\label{eq:Max1}
\end{equation}

\begin{equation}
\frac{\partial \textbf{B}}{\partial t} = -c \nabla \times \textbf{E}
\label{eq:Max2}
\end{equation}

\begin{equation}
\frac{d\rho_{e}}{dt} + div \textbf{j}_{e} = 0
\label{eq:Max3}
\end{equation}

\begin{equation}
\frac{d\rho_{i}}{dt} + div \textbf{j}_{i} = 0
\label{eq:Max4}
\end{equation}

%\begin{equation}
%\nabla \cdot \textbf{E} = 4 \pi \rho
%\label{eq:Max3}
%\end{equation}

%\begin{equation}
%\nabla \cdot \textbf{B} = 0
%\label{eq:Max4}
%\end{equation}

\textcolor{black}{Here $\rho_{i}$, $\rho_{e}$, and $\textbf{j}_{e}$, $\textbf{j}_{i}$ are the electron and ion charge densities and corresponding current densities. Typically Eq. \ref{eq:Max4} is neglected in many problems, i.e., assuming the plasma ions are immovable. 
Eq. \ref{eq:Max3} is solved by methods of current weighting \cite{villasenor1992rigorous}. 
In a reference frame moving with velocity $\textbf{v}_{R}$ the ion current density is not zero, it becomes $\textbf{j}_{i} = \textbf{v}_{R} \rho_{i}$. Similarly, the electron current is also $\textbf{j}_{e} = \textbf{v}_{R} \rho_{e}$, and the total current is zero. However, with any charge separation under external forces the effect of plasma currents may become essential.}
\textcolor{black}{One may exclude artificially $v_R$ from the electron part of the current calculation. However, the charge separation effects in this case vanish.}

\begin{figure}
\centering
\includegraphics[scale=0.78]{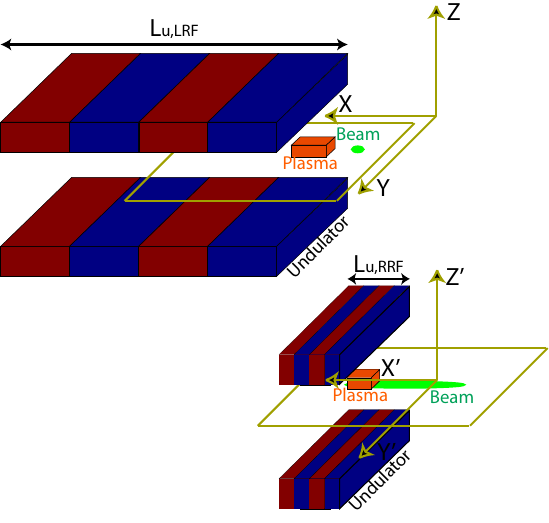}
\caption[XXXX]{Scheme of the space elongation of the Gaussian electron beam (green) and contraction of the undulator (red/blue) and plasma (orange) when going to a relativistic reference frame.
}
\label{PICScheme}
\end{figure}

The simulation of the FEL process via PIC requires of meters of propagation, i.e., the order of magnitude for the electron beam to achieve bunching allowing the start of coherent radiation amplification is of $\approx 1-3$ m (gain length \cite{kondratenko1980generating}). Let's consider a mono-energetic electron beam of $15$ $\mu m$ FWHM length, $\gamma_B = 300$ in the LRF and a RRF $S'$ moving with $\gamma_{R} = 10$. Following Lorentz transformations \cite{lorentz1921deux}:

\begin{equation}
x' = \gamma_{R} (x-v_{R}t)
\label{eq:Lor1}
\end{equation}

\begin{equation}
t' = \gamma_{R} (t-\frac{v_{R}}{c^2}x)
\label{eq:Lor2}
\end{equation}

\begin{equation}
v_R = c \sqrt{1-\frac{1}{\gamma_R^2}}
\label{eq:RRFVel}
\end{equation}

%\begin{equation}
%v = c \frac{\sqrt{\gamma^2-1}}{\gamma}
%\label{eq:RRFVel}
%\end{equation}
  
with the prime denoting the RRF, $v_{R}$ the longitudinal frame speed and c the speed of light. One finds that in $S'$, the undulator length is reduced by $\frac{L_{und}}{\gamma_R}$ (figure \ref{PICScheme}). However, for a Gaussian initial electron beam, the transformation of the electron distribution using Eqs. \ref{eq:Lor1}, \ref{eq:Lor2} gives the following elongation of the beam: 

%\begin{equation}
%e^{ -\frac{ (x+u_{b}t)^2 }{ \sigma_{x}^{2} } } = e^{ -\frac{ \gamma^{2}_{r} }{ \sigma_{x}^{2} } [x'(1-\frac{u_{b}v_{r}}{c^2}) + t'(u_{b}-v_{r})]^2 } = e^{ -\frac{ \gamma^{2}_{r} }{ \sigma_{x}^{2} } (1-\frac{u_{b}v_{r}}{c^2})^{2} [x' + t'(1-\frac{u_{b}v_{r}}{c^2})^{-1} (u_{b}-v_{r})]^2 }  
%\label{eq:BGauss}
%\end{equation}

\begin{equation}
e^{ -\frac{ (x+u_{b}t)^2 }{ \sigma_{x}^{2} } } = e^{ -\frac{ \gamma^{2}_{R} }{ \sigma_{x}^{2} } (1-\frac{u_{b}v_{R}}{c^2})^{2} [x' + t'(1-\frac{u_{b}v_{R}}{c^2})^{-1} (u_{b}-v_{R})]^2 }  
\label{eq:BGauss}
\end{equation}

with $u_{b}$ the beam longitudinal velocity, $v_{R}$ the RRF velocity and $\sigma_{x}$ the beam size on the LRF. Considering the distribution as done at the initial time of the RRF ($t'=0$) and the relation \ref{eq:RRFVel}, the beam size becomes in the RRF: 

\begin{equation}
\sigma_{x}' = \sigma_{x}\frac{\gamma_B}{\gamma_B \gamma_R - \sqrt{\gamma^2_B-1}\sqrt{\gamma^2_R-1}}
\label{eq:Belon}
\end{equation}

with $\gamma_B$ the electron beam gamma. Therefore, the aforementioned electron beam becomes $\approx 0.3$ mm in such RRF.
These values permit to perform a reasonable PIC simulation of the entire process.
As it is to be expected, the changes of lengths also affect the resolution needed. 
\textcolor{black}{Since the beam propagation length in plasma and/or undulator is $L/\gamma_R$ (with L being $L_{undulator}$ or $L_{plasma}$) and plasma propagation length through the beam is $l_b \gamma_R$ the optimal $\gamma_R$ is $\sqrt{L/l_B}$. However, if computer resources allow it, $\gamma_R$ can be reduced.}

%\textcolor{black}{From a simulation time point of view, the optimal $\gamma_R$, should be equal to the largest between $\sqrt{L_u / l_b}$ or $\sqrt{L_{plasma} / l_b}$, i.e., shortest simulation time needed in RRF. However, in occasions such $\gamma_R$ could need high computer resources, thus. a different $\gamma_R$ is preferred.}
%\textcolor{black}{Since we have at least two objects, the optimal $\gamma_R$ equals to $\sqrt{L_u / l_b}$ or $\sqrt{L_{plasma} / l_b}$. However, this value should be correctly chosen depending on the computer ressources.} 
Considering the FEL simulation case, the smaller feature that has to be resolved is, in the LRF, the radiation wavelength given by \cite{pellegrini2007lasers}:

\begin{equation}
\lambda_l = \frac{\lambda_u}{2\gamma^2_B} (1+ \frac{K^2_u}{2})
\label{eq:RadLambda}
\end{equation}

with $\lambda_l$ the radiation length, $\lambda_u$ the undulator period and $k_u$ the deflection parameter of the undulator. For a 1 T peak field and a 2 cm period undulator for our example electron beam one obtains $\lambda_l\approx 304$ nm.  
However, as the simulation is done in a RRF, the RRF $\lambda_l$ adds a dependence in $\gamma_R$ as follows \cite{doppler1842coloured}:

\begin{equation}
\lambda'_l = \lambda_l \sqrt{\frac{1+\beta_R}{1-\beta_R}} \approx \frac{\lambda_{u}}{\gamma^{2}_{B}} \sqrt{\frac{1+\beta_R}{1-\beta_R}}
\label{eq:Uelon}
\end{equation}

\begin{equation}
\beta_R = \sqrt{1-\frac{1}{\gamma_R^2}} 
\label{eq:beta}
\end{equation}

%\begin{equation}
%\lambda'_l = \lambda_l \gamma_r = \frac{\lambda_{u}}{\gamma^{2}_{B}}\gamma_r 
%\label{eq:Nnormi}
%\end{equation}

Thus, changing the required resolution on the LRF from nm to $\mu$m for $\gamma_R > 10$, which is a reasonable spatial resolution for PIC, e.g, in our example $\lambda'_l\approx6$ $\mu m$. While it may be tempting to increase $\gamma_R$ as much as possible, this will cause the electron beam to increase in length, therefore requiring a longer simulation window and number of cells to keep the $\mu$m resolution. A badly chosen simulation $\gamma_R$ can increase too much the computational resources needed. The interplay between the different element sizes and their implication on the resolution and resources has to be carefully considered case by case.
The resolution requirements can be summarized with a condition on the relativistic resolution factor $R_{RF}$:

\begin{equation}
R_{RF} = \frac{8 L_{wind}}{N_x} \sqrt{\frac{1-\beta_R}{1+\beta_R}} < \lambda_l 
\label{eq:Rrf1}
\end{equation}

with $N_x$ the number of simulation cells in the longitudinal (propagation) direction and $L_{wind}$ the total longitudinal window size in the RRF. \textcolor{black}{Furthermore, using Eq. \ref{eq:RadLambda} and simplifying $(1+ \frac{K^2_u}{2}) \approx 1$ (for undulator $K^2_u<1$) one can obtain:}

\begin{equation}
\textcolor{black}{R_{RF} \approx \frac{16 \gamma^2_{B} L_{wind}}{N_x} \sqrt{\frac{1-\beta_R}{1+\beta_R}} < \lambda_u}
\label{eq:Rrf2}
\end{equation}

%\begin{equation}
%R_{RF} = \frac{L_{window}}{N_x \gamma_r} < \lambda_l 
%\label{eq:Nnormi}
%\end{equation}

%\begin{equation}
%R_{RF} = \frac{\gamma^2_{B} L_{window}}{N_x \gamma_r} < \lambda_u
%\label{eq:Nnormi}
%\end{equation}

\begin{figure}
\centering
\includegraphics[scale=0.4]{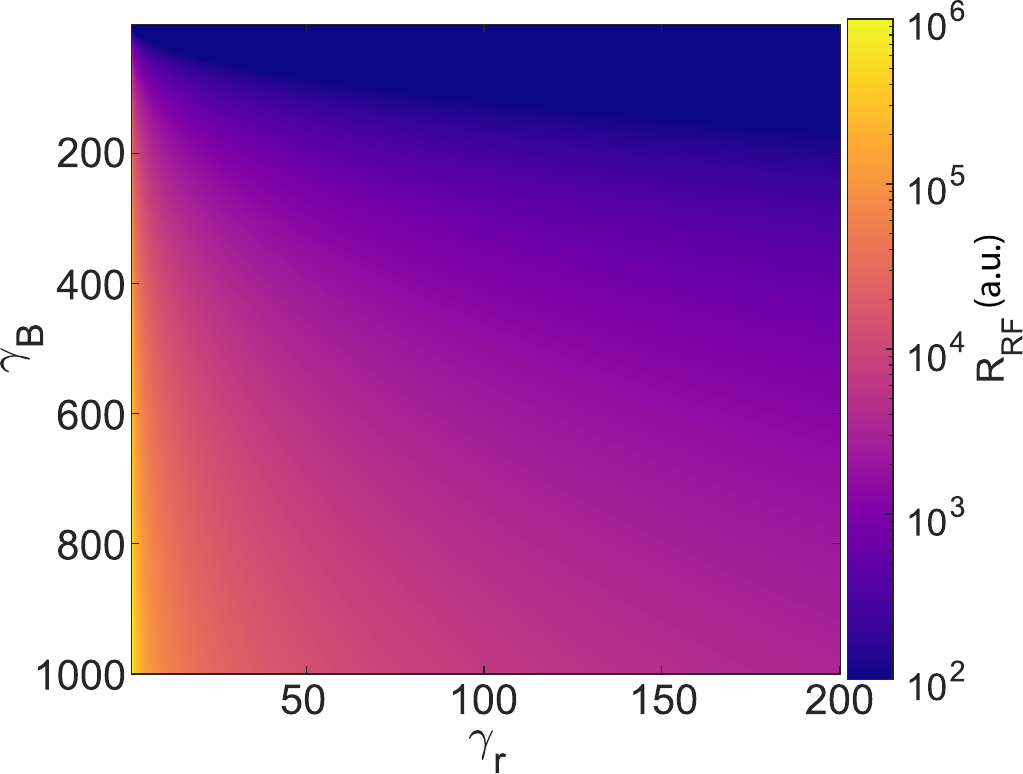}
\caption[XXXX]{ Dependance of the relativistic resolution factor $R_{RF}$ on $\gamma_{B}$ and $\gamma_{R}$ when considering $\frac{16 L_{wind}}{N_x}=1$.}
\label{RRF}
\end{figure}

 The factor 8 of Eq. \ref{eq:Rrf2} comes from the fact that at least 8 points \textcolor{black}{evenly distributed} should be needed to properly determine a sinusoidal wave.
The relativistic resolution factor $R_{RF}$ shows well the interplay of the changes on the different components of the simulation input parameters due to the relativistic nature of the RRF. Figure \ref{RRF} presents the changes in $R_{RF}$ with respect to both $\gamma_B$ and $\gamma_R$.

Regarding the fields, due to the nature of the Maxwell equations, Eqs. \ref{eq:Max1}-\ref{eq:Max4} stay the same. However, the undulator dipolar magnetic field in the LRF is transformed as follows:

\begin{equation}
B'_{\parallel} = B_{\parallel}
\label{eq:M1}
\end{equation}

\begin{equation}
E'_{\parallel} = E_{\parallel}
\label{eq:M2}
\end{equation}

\begin{equation}
B'_{\perp} = \gamma_R (B_{\perp} - \frac{\textbf{u}}{c^2} \times \textbf{E})
\label{eq:M3}
\end{equation}

\begin{equation}
E'_{\perp} = \gamma_R (E_{\perp} + \textbf{u} \times \textbf{B})
\label{eq:M4}
\end{equation}

with $u$ the relative velocity between frames, $\perp$ and $\parallel$ the perpendicular and parallel directions with respect to the RRF propagation direction. 
%Therefore, an electric field of $E'_{\perp} = \gamma_r u \times \textbf{B}$ in the RRF.

When adding a plasma the same length contraction occurs as with the undulator. However, in the case of the plasma, as a consequence, the density ($N_e$) in the RRF is higher:

\begin{equation}
N'_e = N_e \gamma_R
\label{eq:Nnormi}
\end{equation}

Thus, depending on the chosen $\gamma_R$, if the plasma is too dense and thin a higher longitudinal resolution $\Delta X$ may be needed. Also to make sure that the plasma effect is well resolved for such cases the number of macroparticles per cell for the plasma may also be increased.

\textcolor{black}{The change of the plasma density rises the question about the plasma wavelength and wake size. Even though the wake behavior in plasma is out of scope of FEL topic, because the wake and its stability do not affect characteristics of beams, it is interesting to understand possible further plasma evolution.  
Since the plasma electron density and their mass increase with $\gamma_R$ linearly, the plasma frequency does not depend on the reference frame. Therefore $\lambda_p$ should be also constant. However just behind a driver, there is a wave moving with the driver group velocity, $v_{gr}$, and the density is a running wave $n_{e}=n(\frac{x+v_{gr}t}{\lambda_p})$ in LRF.
After the Lorentz transformation, the length of this wave increases $\gamma_R$ times, $\lambda'_p =\lambda_p \gamma_R$ in RRF. Nevertheless, after the driver passes through plasma the running wave vanishes and $\lambda'_p$ should become independent on $\gamma_R$, with a peak density distribution inside the wave period depending on $\gamma_R$.
 In our test simulations we observed the increase of wake length in first periods with $\gamma_R$, while the length of post density perturbation shows the length not increasing with $\gamma_R$. Nevertheless, the problem, which is not trivial, requires special consideration that is out of the scope of this paper.}

All static elements in the LRF (e.g., undulator, plasma), move in the opposite direction to the RRF propagation with a speed of:

\begin{equation}
V'_{element} = -c \sqrt{1-\frac{1}{\gamma^2_{R}}} = -c \beta_R
\label{eq:Vprime}
\end{equation}

For the particle push, the leap-frog scheme \cite{birdsall2018plasma} is here used, providing a second-order accuracy, a good compromise between exactitude and calculation speed.

%\begin{equation}
%\rho(x) = N q_i w_{i(x-x_i)}
%\label{eq:Nnormi}
%\end{equation}

%where $N$ is the number of particles, $q_i$ is the charge of the $i$th particle, $\mathbf{x}_i$ is its position, and $w_{i(\mathbf{x}-\mathbf{x}_i)}$ is the weight function for the $i$th particle.

\textcolor{black}{The simulation uses a moving window carefully elaborated to work for any $\gamma_R$, automatically adjusting its movement to be well adequate to the RRF time-step.} The boundaries of the problem are extremely important to avoid any artificial interference with the undulator and radiation fields and prevent particles from reflecting back into the simulation domain after reaching the edge of the grid. In this particular case, the fields are absorbed upon interaction with the boundaries. While the electron beam macroparticles become dummy particles when going beyond the boundaries. \textcolor{black}{For all simulations in this work the number of particles per cell is 4.} \textcolor{black}{The ion current is directly calculated (essential for plasma in the RRF) in this code. In addition, the approximation of using the electron current as an analogue of the ions motion was tried. Results of both calculations were different, which implies that the ion current effect due to charge separation was essential. Therefore, the \textcolor{black}{proper} direct method was chosen.}
\textcolor{black}{The particle populations (ion, electron, beam electrons) are split. The plasma ions and electrons use the same weighting, otherwise, the discrepancies cause instabilities.}

Regarding the plasma, as mentioned before, in the RRF it moves opposite to the electron beam and window, therefore, it will always enter the window from one side an disappear through the other boundary.

\textcolor{black}{The code itself started as a modified version of the extensively used FPLaser PIC code. However, the multiple challenges to adapt it for RRF made it clear that the changes required were numerous enough to justify the making of a new code. FPLaser has been extensively tested and compared with experiment and other codes for years. This relativistic code, when using $\gamma_R=1$, gives similar results as the non-relativistic frame FPLaser and agrees well with other codes (e.g., FBPIC \cite{lehe2017fourier}, OSIRIS \cite{fonseca2002osiris}, WARP \cite{friedman1990warp}, PIConGPU \cite{burau2010picongpu}), which indicates that electromagnetic field calculation, current calculation, particle weighting, etc are done correctly.}

\section{Understanding the relativistic reference frame }

One of the difficulties for data treatment and comparisons between RRF and LRF that arise from the Lorentz transformations (Eqs. \ref{eq:Lor1}, \ref{eq:Lor2}) of the particles is the time difference. From equations \ref{eq:Lor1}, \ref{eq:Lor2}, it can be seen that as the electron evolves in the RRF, its momentum and position changes, an therefore, the corresponding LRF $t$ to  its $t'$. While straightforward for a single electron, when considering two electrons one finds that for an instant $t'$, if the electrons have different position and momentum, their respective $t_{e1}$ and $t_{e2}$ are not necessarily equal, thus, the instant $t'$ translates into a span of time (Figure \ref{MultiRRFs2LRF}): 

\begin{equation}
\Delta T_{e1,e2} = \frac{\gamma_R}{c^2} v_{R} ( x'_{e1} - x'_{e2})
\label{eq:delta2electrons}
\end{equation}

The quantity of simultaneous electrons in a beam only aggravates this issue.
Therefore, it is difficult to judge the exact state in the LRF of the full beam, especially during interaction with other elements, e.g., plasma. 
Two solutions present themselves to this conundrum: 

\begin{figure}
\centering
\includegraphics[scale=0.7]{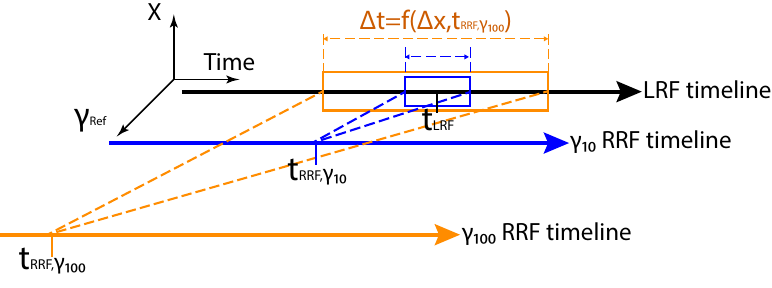}
\caption[XXXX]{Visual representation of the Lorentz transformation towards the LRF of a group of electrons with different $p_x$ and $x$ at an instant $t_{RRF,\gamma10}$ ($t_{RRF,\gamma100}$) in a RRF of $\gamma_R=10$ ($\gamma_R=100$). 
 }
\label{MultiRRFs2LRF}
\end{figure}

%One of the difficulties for data treatment and comparisons between RRF and LRF that arise from the Lorentz transformations (Eqs. \ref{eq:Lor1}, \ref{eq:Lor2}) of the particles is the time difference. From equations \ref{eq:Lor1}, \ref{eq:Lor2}, it can be seen that as the electrons evolve in the system, their $\gamma_B$ changes, so, in an instant $t'$, each electron will have a different $t$ after going back to LRF. Therefore, it is difficult to judge the exact state in the LRF of the full beam, especially during interaction with other elements. Two solutions present themselves to this conundrum: 

For a group of particles of similar momentum in the RRF propagation direction, one can select $\gamma_R$ and $\Delta X^{'}$ so that the $\Delta T$ after transformation to the LRF is small enough to be negligible or at least acceptable:

\begin{equation}
\Delta T = \frac{\gamma_R \beta_R}{c^2} \Delta X^{'}  = \frac{ \sqrt{\gamma_R^2-1}}{c^2} \Delta X^{'} 
\label{eq:DT}
\end{equation}

However, the transformation from RRF to LRF carries always an error in the quantities of around the order of its variation during $\Delta T$. 

\textcolor{black}{Due to the time quantization inside the simulation, the evolution of $\Delta T$ corresponding to a step in $t'$ depends on both the simulation time step and the system particles momentum in the RRF propagation direction. Therefore, if the momentum in the RRF propagation direction for certain particles is significantly larger than others the $\Delta T$ could vary substantially between time steps. 
Furthermore, it may be impossible to directly reconstruct a specific time in the LRF even with full information of all time steps during the RRF simulation. With the only direct solution being using a huge time resolution in the RRF which could counter the simulation speed gains obtained over the LRF.
That is why the condition of similar momentum particles was assumed for Eq. \ref{eq:DT}.}
Once each group dynamics have been transformed to the LRF, the time evolution in the LRF should be able to be reconstructed for a good resolution simulation. In particular cases, this LRF reconstruction process can be easier and simpler if the change in momentum is gradual and/or periodic, e.g., undulator field.
Furthermore, for two different $\gamma_R$ RRFs, let's say 10 and 100, the time window produced by an instant in their respective reference frames will differ of a factor 10 (Figure \ref{MultiRRFs2LRF}), for a same $(\Delta X^{'} , \Delta p^{'}_x)$. So, even between RRFs the comparison is not necessarily intuitive. 
% \textcolor{green}{Due to the time quantization inside the simulation, one can only know the state of the particles at each time step, therefore, if the momentum in the RRF propagation direction for certain particles is significantly larger than others, one could end up without enough information to reduce the $\Delta T$ of the entire system. Furthermore, it may be impossible to directly reconstruct a specific time in the LRF even with full information of all time steps during the RRF simulation. With the only direct solution being using a huge time resolution in the RRF which could counter the simulation speed gains obtained over the LRF.
% That is why the condition of similar momentum particles was assumed for Eq. \ref{eq:Nnormi}.}

% \textcolor{blue}{The condition of small momentum variations comes from the fact that in the simulation one can only know the state of the particles at each time step, and if the momentum in the RRF propagation direction for certain is significantly larger than others, one could end up with not enough information to reduce the $\Delta T$ of the entire system.}

% \textcolor{blue}{Nevertheless, the simulations are often used to understand non-linear phenomena, thus, the condition of $\Delta p^{'}$ small for a group of particles is not usually achievable when considering an entire beam. In such cases, during the post-treatment of data, one can divide the number of particles into groups based on small enough $(\Delta X^{'} , \Delta p^{'}_x)$ (assuming $x$ as the RRF propagation direction) so that, $\Delta T$ for the is acceptable.} 

The issue with the previous solution is the amount of post-processing necessary to do it albeit being technically possible. The second solution is to base the analysis to only invariants. For example, the transverse dynamics are invariant between reference frames, radiated energy, beam charge in vacuum. 

\textcolor{black}{In addition, the use of $\gamma_R = \gamma_B$ difficults the spectral analysis as the beam is static in the RRF and the emission is in $2\pi$. On top of this, for such high $\gamma_B$ beams, $\gamma_R = \gamma_B$ reduces considerably the undulator period and increases the FEL wavelength thus, having a bunching saturation wavelength much larger than an undulator period. Finally, the beam size increases substantially and the plasma is shortened causing sometimes the need for too high resolution. Therefore, for such studies $\gamma_R = \gamma_B$ does not seem to be appropriate.}

%\textcolor{black}{In addition, the use of $\gamma_R = %\gamma_B$ has no meaning due to the spectral analysis %being unable to differentiate types of beam %modulations.}

\textcolor{black}{These differences between reference frames also affect the understanding of the wakes generated by a driver in an RRF as already alluded previously. 
%The density perturbation (due to a driver) lengths change differently depending on the direction ($\lambda'_{p,\perp}$ and $\lambda'_{p,\parallel}$), and on top of that 
For a fixed time in the RRF the plasma particles are distributed in a $\Delta t_{LRF}$. So, when observing a wake in the RRF, one has to keep in mind that it does not correspond to any particular wake that may appear in the LRF, thus, being a different object, a wake in space-time and not only space. Again, such counter-intuitive subtleties makes even more important the need for a careful post-processing of the data.}

\section{Electron propagation in vacuum and under external field}

Our main motivation for the development of the RRF PIC code is to be able to simulate the electron beam dynamics during propagation inside long plasmas (10s of cms) for their use as pre-buncher, optics and also for the research of plasma undulators. Therefore, the most appropriate benchmarks are the propagation of electron beams in vacuum, in an external field (e.g., undulator) and inside a long plasma. \textcolor{black}{All simulations are done in 2D as the main phenomena that want to be observed would not change significantly from a 3D simulation.}

\begin{table*}
    \centering
    \begin{tabular}{c|c|c|c|c|c|c|c|c|c|c|c|c}
        Case & $\gamma_r$ & $\gamma_b$ & $L_{wind,X}$ & $L_{wind,Y}$ & $dX_{RRF}$ & $dY_{RRF}$ & $D_{plasma,LRF}$ & $L_{plasma,LRF}$ & $L_{plasma,RRF}$ & $\lambda_{u,LRF}$ & $\lambda_{u,RRF}$ & $B_{peak}$  \\
        &  &  & $mm$ & $mm$ & $\mu m$ & $\mu m$ & $cm^{-3}$ & $cm$ & $mm$ & $cm$ & $mm$ &  $T$ \\
        V. & 50 & 300 & 7 & 2  & 0.5 & 0.65 & --- & --- & --- & --- & --- & --- \\
        Und.   & 50 & 300  & 7 & 2  & 0.5 & 0.6 & --- & --- & --- & 2 & 0.4 & 1 \\
        Pl. & 10 & 300 & 1.4 & 0.7  & 0.107 & 0.359 & $3\times10^{15}$ & 20 & 20 & --- & --- & --- \\
        Pl.2 & 25 & 1200 & 7.5 & 1.4  & 0.21 & 0.9 & $3\times10^{16}$ & 20 & 8 & --- & --- & --- \\
        Pl.+Und. & 10 & 300 & 1.4 & 0.7  & 0.11 & 0.36 & $3\times10^{15}$ & 20 & 20 & 2 & 2 & 1 \\    \end{tabular}
    \caption{Simulation parameters for the vacuum (V.), beam inside undulator (Und.), beam inside plasma (Pl.) and beam+plasma+undulator (Pl.+Und.) cases. $dX_{RRF}$ ($dY_{RRF}$) resolution in the propagation (transverse) direction, $D_{plasma,LRF}$ plasma density in the LRF, $L_{plasma,LRF}$ ($L_{plasma,RRF}$) plasma length in the LRF (RRF), $\lambda_{u,LRF}$ ($\lambda_{u,RRF}$) undulator period in the LRF (RRF), $B_{peak}$ undulator peak field.} 
    \label{tab:param}
\end{table*}

% For the following in-vacuum simulation the parameters in Table \ref{tab:param} has been used. 

The simulations parameters are presented in Table \ref{tab:param}. 
%The simulations have been performed with a total window size of XRL and FLNG in the longitudinal (vertical) with N particles per cell. Two $\gamma_r$ cases are compared in vacuum, 10 and 100, with MPX (MPY*cores) cells and MPX (MPY*cores) cells respectively, to conserve the same resolution $\Delta X$ and $\Delta Y$ (Table \ref{tab:param}). 
In all cases a beam of 30 pC, 40 $\mu m$ transverse diameter and 15 $\mu m$ length in the LRF has been used. After definition in the LRF the code itself calculates the equivalence to the RRF with the chosen $\gamma_R$ following the prior equations. 
As previously mentioned, the electron beam length is increased (Eq. \ref{eq:Belon}), however, the same total charge is kept constant, thus, making the charge density lower in the RRF. For all simulations an initial transverse momentum has been applied to the beam. 
%$P_y = 2.5\cdot10^{-4} P_x$ is used for the initial electron beam. 
The time and space are normalized by $\omega$ and $c/\omega$ respectively, with $c$ the speed of light and $\omega=\omega_{plasma,cgs}=2e \sqrt{\frac{\pi N_e \gamma_R}{m_e}}$ (with $m_e$ the electron mass and $e$ the electron charge), except in the case where $\omega_{plasma,cgs}$ is lower than $\frac{2 \pi \gamma_R c}{\lambda_u}$, in that case the latter is used instead for convenience.

\begin{figure}
\centering
\includegraphics[scale=0.9]{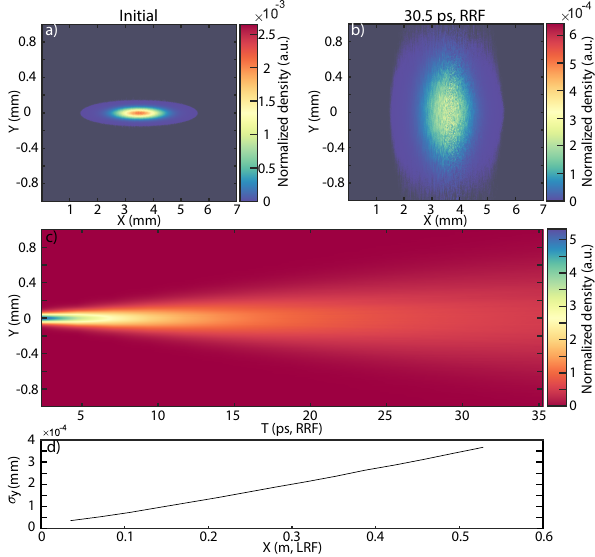}
\caption[XXXX]{Electron beam density distribution (after a transverse smoothing of 5 px) at a) the start of the simulation and b) after 30.5 ps RRF ($\approx0.45$ m in LRF) (Table \ref{tab:param} V.). c) evolution in time of the sum along the longitudinal axis of the beam density distribution. d) Transverse size RMS evolution. Propagation direction to the left.}
\label{FigVacuum}
\end{figure}

In-vacuum the only forces on the electron beam are those created by its own electrons electric field and current. In order to avoid any non-physical currents due to the sudden calculation of the relativistic electron beam movement in vacuum during the first steps, an initial background of artificial ions perfectly compensating the electron beam charge is used. During a time $T_{ion}$ the charge of these ions will be damped linearly until their total disappearance. This allows for a better initialization of the fields generated by the electron beam currents. $T_{ion}$ is here set as $\approx0.1$ ps in the RRF.
\textcolor{black}{In addition, a very small transverse divergence (much smaller than that in realistic cases) is added to simulate a realistic case and also to avoid any non-physical self-focusing of the electron beam due to Cherenkov instabilities \cite{pukhov2020x}. The introduction of a transverse momentum reduces the normalized momentum, i.e., $\frac{p_x}{p_{total}}$, thus, avoiding the numerical error that may allow electrons to go above the speed of light.}
Figure \ref{FigVacuum} shows the propagation of an electron beam with initial $\sigma^,_{y,RMS} = 0.5$ mrad (root mean squared) transverse divergence in vacuum during $\approx35.5$ ps, which corresponds to $\approx0.55$ m in the LRF, with the simulation parameters in Table \ref{tab:param} V.. 
Figure \ref{FigVacuum}a,b presents the initial beam density distribution and after 30.5 ps RRF of propagation and Figure \ref{FigVacuum}c the transverse density projection evolution at each time step. It is seen that the beam properly diverges as expected. \textcolor{black}{Only taking into account the initial beam divergence, the beam after $\approx 0.55$ m is expected to have $\sigma_y=315$ $\mu m$, the simulation gives 364 $\mu m$. On top of that the size evolution (Figure \ref{FigVacuum}d) is slightly non-linear. This is due to the effect of the space charge that, as expected, increases the initial divergence of the beam non-linearly, approaching $\sigma^,_{y,RMS} \approx 0.75$ mrad at 0.55 m. Nevertheless an exact quantitative divergence evolution in vacuum due to the space-charge requires of a 3D PIC simulation. Still this shows that a high resolution and accuracy simulation of vacuum propagation along $\approx0.55$ m taking into account the intra-beam forces, can be done in just 14 real time hours and 220 cores, which is already not possible in normal PIC codes. }

%The divergence increases over time (Figure \ref{FigVacuum}d) due to the high initial charge density going from $\sigma^,_{y,RMS} = 0.5$ mrad to $\sigma^,_{y,RMS} \approx 0.78$ mrad. Therefore, 

%a high resolution simulation of vacuum propagation along $\approx0.55$ m has been done in just 14 real time hours and 220 cores, which is already not possible in normal PIC codes.  

\begin{figure}
\centering
\includegraphics[scale=0.7]{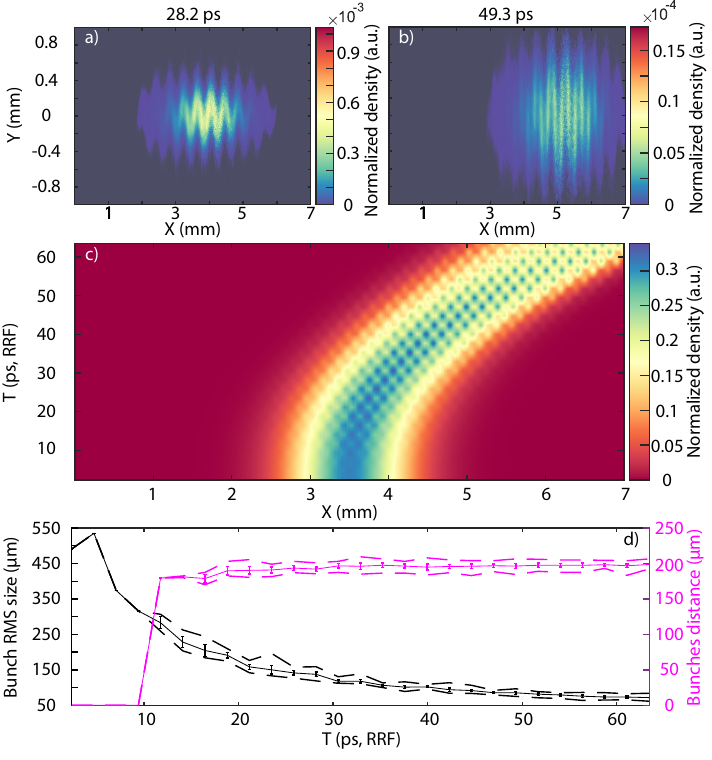}
\caption[XXXX]{Electron beam density distribution (after a transverse smoothing of 5 px) at times a) 28.2 ps ($\approx0.42$ m in LRF) and b) 49.3 ps ($\approx0.74$ m in LRF) in RRF (Table \ref{tab:param} Und.). c) Evolution in time of the Density distribution sum along the transverse axis. Electron beam d) inter-bunch distance and bunch RMS size mean (solid line), std (error bar), maximum and minimum (dotted line). Propagation direction to the left.}
\label{FigUND}
\end{figure}

Figure \ref{FigUND} presents a $\approx1$ m propagation of the electron beam inside an undulator (Parameters at Table \ref{tab:param} Und.). 
Special care has to be taken with the start of the undulator field in the simulation. The field cannot start suddenly as the electrons would jump from a zone without field to one with a strong enough field in a single step, causing a kick in the transverse direction. To solve it, a gentle slope is used, which has been found to erase this effect. The limitation comes from the leap frog scheme accuracy. This quick solution allows to prevent the non-physical kick with minimum effect on the beam dynamics and avoiding the need of unnecessary high resolution.
Inside a periodic undulator field the electron beam starts to oscillate (Figure \ref{FigUND}a) and eventually, due to its own charge and emitted radiation the density starts to acquire a longitudinal modulation (Figure \ref{FigUND}b) \cite{bonifacio1990large}. Following the evolution of the transversely integrated beam density (Figure \ref{FigUND}c), the slow apparition of the density bunches can be clearly seen. Initially, the electron beam starts its oscillation creating a density wave of wavelength 400 $\mu m$, i.e., equal to the undulator $\lambda^{'}_{u}$ (RRF).
After around 8 ps in the RRF, the bunches start to appear (Figure \ref{FigUND}d), with an inter-bunch (\textcolor{black}{i.e., distance between the distribution peaks of two consecutive bunches}) distance of 200 $\mu m$ corresponding to $\lambda^{'}_{u} /2$. Following the 3D phenomenological equations for FEL obtained by Ming Xie \cite{xie1995design}, for a self-amplified-spontaneous-emission (SASE \cite{kim1986analysis}) FEL, the gain ($L_{g,3D}$) and saturation lengths ($L_{s,3D}$) for our configuration are $\approx 6.7$ cm ($\approx 6$ ps in RRF) and $\approx 1.22$ m ($\approx 80$ ps in RRF) respectively. 
\textcolor{black}{In the simulation, the beam starts to show clear differentiable bunches after $\approx 8$ ps, which agrees with the equations prediction ($L_{g,3D} \approx 6.7$ cm or $6$ ps in RRF)}. By the end of the simulation, i.e., $\approx 63$ ps, the mean bunch size (i.e., longitudinal size of single bunch inside the electron beam) starts to saturate, thus, it is close enough to the Ming Xie equations results for saturation ($\approx 80$ ps in RRF). The saturation mean bunch size is of around $\sigma_{x,bunch}=60$ $\mu m$ RMS (Figure \ref{FigUND}d). \textcolor{black}{The slight differences are easily explained by the phenomenological character of the equations and the simulation being done in 2D.}
A proper longitudinal density modulation can be observed in Figure \ref{FigUND}c.

With time, due to the periodical change of longitudinal momentum of the electrons during their transverse oscillation, the beam propagates slightly slower than the simulation window causing its drift to the back of the window (Figure \ref{FigUND}b,c). However, this can be easily solved by tweaking the window velocity.

%\textcolor{black}{A comparison of the same simulation with different $\gamma_r$ is shown in Figure \ref{FigUNDComp}. The change of $\gamma_r$ implies a change in the undulator period (Eq. \ref{eq:Uelon}) and beam size  (Eq. \ref{eq:Belon}) in the RRF (Table \ref{tab:param}), thus, making necessary the adjustment of the simulation window parameters.}

\textcolor{black}{Additional simulations were done with the individual elements (plasma, external field, beam) and their combinations for long distances and no numerical instabilities were observed.}

%\section{RRF plasma-beam dynamics}
\section{RRF beam dynamics inside low density plasma}

\begin{figure}
\centering
\includegraphics[scale=0.75]{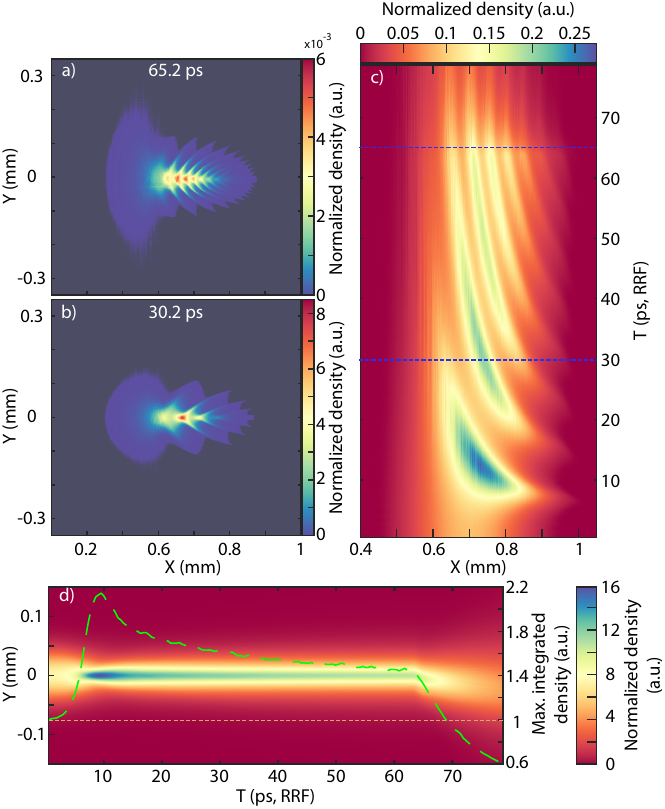}
\caption[XXXX]{Electron beam density distribution (after a transverse smoothing of 5 px) at times a) 65.2 ps ($\approx20$ cm in LRF) and b) 30.2 ps ($\approx9$ cm in LRF) in RRF (Table \ref{tab:param} Pl.). c) Evolution in time of the Density distribution sum along the transverse axis with the distributions times marked with blue dotted lines. d) Evolution in time of the Density distribution sum along the longitudinal axis with the maximum density evolution (green dotted line). Propagation direction to the left.}
\label{FigPlas}
\end{figure}

\begin{figure*}
\centering
\includegraphics[scale=1.05]{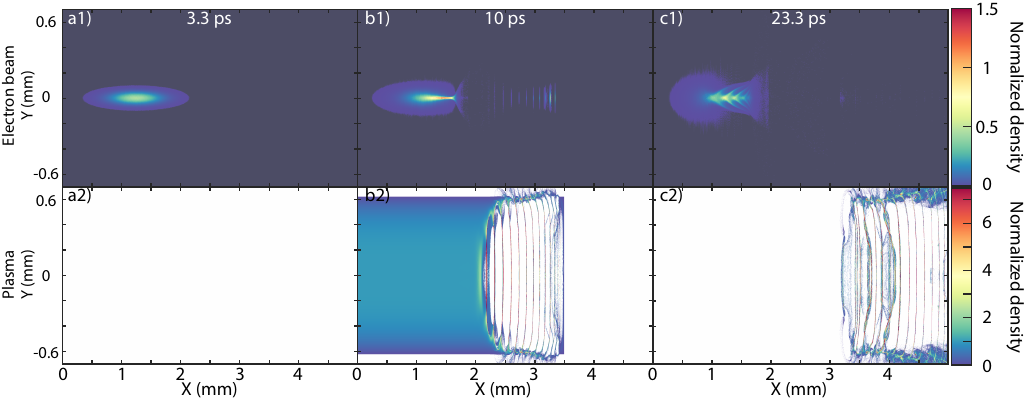}
\caption[XXXX]{a1, b1, c1 Electron beam density distribution (after a transverse smoothing of 5 px) and a2, b2, c2 plasma density distribution after a) 3.3 ps ($\approx2.5$ cm in LRF), b) 10 ps ($\approx7.5$ cm in LRF) and c) 23.3 ps ($\approx17$ cm in LRF) of propagation (Table \ref{tab:param} Pl.2). Propagation direction to the left.}
\label{FigPlas2}
\end{figure*}

As previously mentioned, the propagation of an electron beam inside a cm size plasma can be quite useful for its optics properties \cite{pathak2023focusing, rosenzweig1989beam} and also as a pre-buncher. In Figure \ref{FigPlas}, the results of an 80 ps, in RRF ($\approx 24$ cm in LRF), simulation of a beam propagation through a low density 20 cm plasma (Table \ref{tab:param} Pl.) are displayed.
Once inside the plasma, the head of the Gaussian beam generates a wakefield strong enough to cause a localized focusing on the back of the beam (Bunch 1). At the same time, Bunch 1 interacts with the wakefield and propagates through the beam towards the head (Figure \ref{FigPlas}c) as the front part slightly defocuses, i.e., diverges (Figure \ref{FigPlas}a,b). Concurrently, when the Bunch 1 achieves enough local density (Figure \ref{FigPlas}c), it creates a wakefield that focuses a posterior part of the beam (Bunch 2) while diverging in a half-moon shape.
This bunching process due to the plasma that continues during the entire propagation is well illustrated in Figure \ref{FigPlas}c. The time between the apparition of a bunch and the creation of the next one takes $\approx 7.7$ ps ($std=0.7$ ps), which is $\approx 2$ cm in the LRF. By the end of the 20 cm of plasma, a total of 11 bunches separated by $47.7\pm12.5$ $\mu m$ can be observed inside the beam (Figure \ref{FigPlas}a,c), i.e., the back bunches are closer together than the front ones. 
In addition, even though a fishbone structure appears, transversely, the beam is focused inside the plasma by more than 1.4 times its initial density (Figure \ref{FigPlas}d). Once outside the plasma, the beam starts to diverge while keeping its structure. 

%\textcolor{black}{Same parameter simulations have been carried with different $\gamma_R$, including $\gamma_R$=1 (though, less propagation due to being time consuming), and all of them show the fishbone structure due to the plasma.}

\textcolor{black}{Same parameter simulations have been carried with different $\gamma_R$, and all of them show the fishbone structure due to the plasma. For $\gamma_R$=1 (less propagation was simulated due to being time consuming) we observed beam focusing and modulation only in plasma. With a $\gamma_R$ increase the modulation structure does not change exhibiting the exact relativistic increase between the
density peaks, while the focusing (transverse phenomena) is the same as expected.}

The code is also capable of higher energies and densities. Figure \ref{FigPlas2} presents the case of a 600 MeV beam in a $3\times10^{16}$ $cm^{-3}$ plasma (Table \ref{tab:param} Pl.2). The initial Gaussian beam (Figure \ref{FigPlas2}a) enters the plasma and as in Figure \ref{FigPlas} starts to be focused but in this case the focusing is strong enough to expel electrons of the wake creating a plasma modulation (Figure \ref{FigPlas2}b). Once out of the plasma, only the front half of the electron beam continues to propagate with a partial longitudinal density modulation while the rest stays trapped in the strongly modulated plasma (Figure \ref{FigPlas2}c). \textcolor{black}{For a previously prepared electron beam (longitudinally energy sorted beam, e.g., magnetic chicane \cite{loulergue2015beam}), this effect could be used as a low or high energies filter, reducing the beam energy spread.}

The full high resolution simulation shown in Figure \ref{FigPlas} (Figure \ref{FigPlas2}) was made in only 50 (30) hours with 150 (210) cores and less than 200 gb of RAM, e.g., a single work station, allowing the research of 10s of cm simulation with a plasma. \textcolor{black}{These simulation already present a great potential for the swift beam bunching using a plasma, however,} An in depth exploration to properly understand the use of plasma as pre-buncher using this code \cite{EPHEMER} is in progress and will be presented in another work as it is out of the scope of this article.

\section{Three elements long distance simulation}

\begin{figure}
\centering
\includegraphics[scale=1.1]{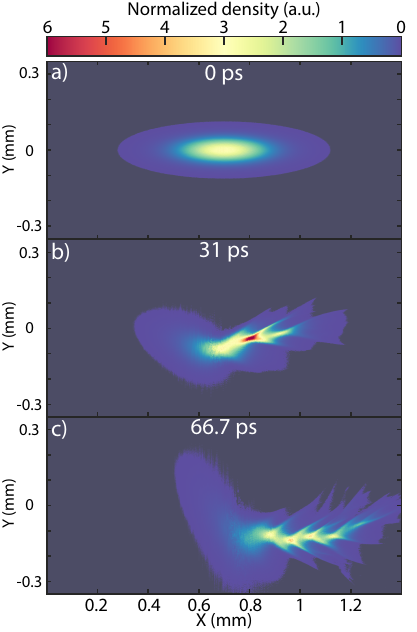}
\caption[XXXX]{Electron beam density distribution (after a transverse smoothing of 5 px) at a) the start and after b) 30.2 ps ($\approx0.09$ m in LRF) (inside the plasma and undulator) and c) 66.7 ps ($\approx0.20$ m in LRF)  (out of the plasma but inside the undulator) of propagation in RRF (Table \ref{tab:param} Pl.+Und.). Propagation direction to the left.}
\label{FigPlasUnd}
\end{figure}

As a last demonstration of the capabilities of the code, Figure \ref{FigPlasUnd} presents the case of an electron beam propagating inside a 20 cm plasma which is at the same time inside a 1 T and 2 cm period undulator (Table \ref{tab:param} Pl.+Und.). The electron beam enters the undulator field slightly earlier than the plasma, thus, starting its oscillation (Figure \ref{FigPlasUnd}a). Once inside the plasma (Figure \ref{FigPlasUnd}b), its modulation due to the wakefield starts as in Figure \ref{FigPlas}. However, due to the oscillation, the wakefield does move in the transverse direction, which imprints a shift in the plasma wakefield induced bunches (Figure \ref{FigPlasUnd}c). Again, an in-depth study of this phenomena is under progress.   

%\textcolor{black}{talk here about the difficulty to imagine a ring of thin plasma going through the e beam ?}

%For a high $\gamma_r$, the plasma will shrink to a length smaller than the electron beam causing a difficult to imagine situation. In the LRF the electron beam can find itself in one of two cases, i.e., part inside the plasma and the rest outside (entrance and exit) or fully inside). However, for a high enough $\gamma_r$, in the RRF at no point the beam will be fully inside the plasma, thus, one can observe a situation in which the there are simultaneously electrons that already exited the plasma, never entered the plasma and are inside the plasma.

\section{Discussion and Conclusion}

In this work, a new PIC \textcolor{black}{code} that uses the advantages of space-time dilation to allow the simulation of complex physical systems for long distances (up to meters) in Cartesian PIC using even a single work station has been presented.
%In this work, the 2D VENTUS code has been presented as a new PIC paradigm that uses the advantages of space-time dilation to allow the simulation of complex physical systems for long distances (up to meters) in Cartesian PIC using even a single work station. 
The non-intuitive aspects of working in a relativistic reference frame (e.g., Gaussian beam elongation, fields transformation, relativistic Doppler) have been discussed. Due to the intrinsic features of the Lorenz transformation (Eqs. \ref{eq:Lor1},\ref{eq:Lor2}), an instant $t'$ in the RRF can result in a time window of $\Delta t$ in the LRF, therefore, making the interpretation of the code raw results non-trivial in several instances. Nevertheless, multiple solutions for the results post-processing have been discussed, however, the appropriated steps to follow can heavily depend on the explored physics and cannot be all elucidated beforehand.
As an example of the code capacities, we have shown the simulation of an electron beam inside an undulator,  with its longitudinal bunching agreeing with the widely used phenomenological FEL equations \cite{xie1995design}. It also has been shown that the code is an appropriate tool for the study of plasma optics and pre-buncher with excellent results, which needs of 10s of cm of beam plasma interaction. The creation of bunches inside the electron beam due to the plasma wakefield generated by the beam itself show a promising future for such device for uses in FEL and more. The insights about this problematic achieved by this code are quite useful and unique and will be properly explored in future works. Finally, the propagation of a beam inside a plasma and undulator has been simulated showing interesting features on the plasma generated bunches, e.g., the bunches shift in the oscillation direction. The usefulness and capacity to unravel new physics by taking advantage of the relativistic reference frame in PIC have been demonstrated and plenty of new results will be obtained with it in the future. In addition, a 3D version is ready, extending the relativistic reference frame advantages of this code to simulations for 3D phenomena. 
We have to note that all these advantages are applicable only for charge particle beams. Simulations for laser pulses or other radiation sources requires special consideration \textcolor{black}{to be done in general, specially its interaction with overdense plasma.}

%\section*{Acknowledgements}

%This work was funded by the JST-Mirai Program, grant No. jP-MjMI17A1 and was partially supported by the ImPACT R\&D Program of Council for Science, Technology and Innovation (Cabinet Office, Government of Japan). This work is also partially supported by ‘Dynamic Alliance for Open Innovation Bridging Human, Environment and Materials’ from the Ministry of Education, Culture, Sports, Science and Technology of JAPAN (MEXT).
%This research was partially conducted with the supercomputer HPE SGI8600 in the National Institutes for Quantum Science and Technology.
%This work was partially achieved through the use of supercomputer system SQUID at the Cybermedia center at Osaka university.

\section*{Author contributions}

D.O.E and A.Z. developed the PIC code and the theoretical background,
D.O.E performed the simulations, 
D.O.E. analyzed and treated the data.
The article was written by D.O.E. with discussions and corrections of A.Z.. and A.R. 
Additional discussions and support from M.M. and M.T..

\section*{Additional Information}

\subsection*{Acknowledgement}
This work used computational resources of the supercomputer Fugaku provided by RIKEN through the HPCI System Research Project (Project ID: hp240331).

\subsection*{Competing interests}
The authors declare no competing interests.

\subsection*{Data availability}

The datasets used and/or analysed during the current study available from the corresponding author on reasonable request.

%XXXXX

%\blindtext \cite{article-minimal}

\bibliographystyle{apsrev4-1} % Tell bibtex which bibliography style to use
\bibliography{UndRelatSim_ArticleNotes.bib} % Tell bibtex which .bib file to use (this one is some example file in TexLive's file tree)

\end{document}